# On the origin of Stark effect of rotons in He-II and the existence of $p = 0$ condensate

Y. S. Jain[1,]*, L. Chhangte[1], S. Chutia[1,2] and S. Dey[1,3]

[1]Department of Physics, North-Eastern Hill University, Shillong 793 022, India
[2]Department of Physics, St Anthony's College, Shillong 793 001, India
[3]Don-Bosco College of Engineering and Technoogy, Assam Don-Bosco University, Guwahati 781 017, India

**Linear Stark effect of roton transition, experimentally observed through microwave absorption in He-II (superfluid $^4$He) in the presence of varying external electric field, is critically analysed. We find that: (i) The effect cannot be explained in terms of conventional microscopic theory (CMT) of He-II which presumes the existence of $p = 0$ condensate and concludes that $^4$He atoms even at $T = 0$ have random motions and mutual collisions which do not support the basic factor (viz. an ordered arrangement of atomic electric dipoles) needed for its occurrence. (ii) The desired order is concluded, rather, by a non-conventional microscopic theory (NCMT) as an intrinsic property of He-II. Accordingly, all atoms in He-II define a close packed arrangement of their wave packets (CPA-WP) with identically equal nearest neighbour distance ($d$), per particle zero-point energy ($\varepsilon_0 = h^2/8md^2$) and equivalent momentum, $h/2d$. (iii) The CPA-WP prevent atoms from having relative motions and mutual collisions capable of disturbing any order of atomic dipoles. As such the NCMT and the observed Stark effect have strong mutual support; whereas the former concludes CPA-WP necessary for the occurrence of the effect, the latter strengthens the experimental support for the former, which means that the observation does not support the presence of $p = 0$ condensate in He-II.**

**Keywords:** Bosons, microwave absorption, roton transition, Stark-effect.

LIQUID helium-4 (LHe-4), a system of interacting bosons (SIB), has been a subject of extensive research[1–8], for its unique behaviour, such as superfluidity (flow of the fluid without viscosity) and related properties which arise when quantum nature of the atoms dominates its low temperature (LT) behaviour at macroscopic scale. Among all the liquids in nature, only LHe-4 assumes superfluidity at $T < T_\lambda = 2.17$ K, because $^4$He atoms can have the largest ratio of the thermal de Broglie wavelength ($\lambda_T = h/\sqrt{2\pi m k_B T}$, with $h$ being the Planck constant, $k_B$ the Boltzmann constant, $m$ the mass of helium atom) to the inter-atomic separation $d$. The motivation for its studies also lies with the fact that the microscopic understanding of the said behaviour is still unclear.

*For correspondence. (e-mail: profysjain@gmail.com)

In what follows from: (i) a proposal of London[1,9] that the Bose–Einstein condensate (BEC) (also known as $p = 0$ condensate) is responsible for superfluidity and related aspects of He-II (superfluid phase of LHe-4), and (ii) the conclusion of a microscopic model of weakly interacting bosons developed by Bogoliubov[10] that interparticle repulsion pushes a fraction of particles to the states of $p \neq 0$, most people working in the field believe that a SIB in its superfluid phase has depleted value of $p = 0$ condensate. The He-II is reported[7,11] to have a maximum of about 10% atoms in the $p = 0$ state and the BEC state of trapped dilute gases (TDG) about 60% (ref. 4). London's proposal was questioned by Landau[12] as soon as it was advanced. Landau rightly argued that LHe-4 is not a system of non-interacting bosons (SNIB) for which Einstein[13] concluded the existence of BEC below $T = T_{BEC} = (h^2/2\pi m k_B)(N/2.61V)^{2/3}$, where $N$ is the total number of particles and V the volume of the system. The issue was debated by Landau and London for a long time until it was settled in favour of London's proposal with the publication of Bogoliubov's model[7].

The conventional microscopic theories (CMT) of a SIB use single particle basis (SPB) by identifying a single particle as the basic unit of the fluid and by assuming that the particles occupy states of a single particle confined to volume V. However, SPB has a missing link with two important realities of a SIB: (a) particles interact with two-body interactions, indicating that a pair of particles should form the basic unit of the system, and (b) on cooling the system to $T$ at which $\lambda_T$ becomes of the order of $d$, the particles assume a state of their wave super-position for which no particle can be described as independent represented by a plane wave; one at least needs two particles with their representative plane waves to find the wave function resulting from their wave super-position and the way it affects the physical behaviour of particles occupying such a state. This is particularly important because superfluidity is undoubtedly a consequence of the wave nature which assumes dominance at LT. In order to avoid the said missing link, Jain's non-conventional microscopic theory (NCMT)[8], rightly uses pair of particles basis (PPB), and it is for this reason that: (i) an apparently simple difference of SPB and PPB renders significantly different $G$-states (symbolized as $G$-state(SPB) and $G$-state(PPB)) which are depicted by the momentum distribution ($N_p$) of particles in Figure 1 $a$ and $b$ respectively, and the respective position distribution in Figure 1 $c$ and $d$, and (ii) only $G$-state(PPB) helps in finding a clear understanding of superfluidity and related properties of He-II[8].

A critical analysis of $N_p$ (Figure 1 $a$) of $G$-state(SPB) (as discussed briefly in Appendix 1, and in detail in Jain[14]), unequivocally establishes that the corresponding energy $E_0$(SPB) does not assume the least possible value as expected. This remained unnoticed for more than six decades, possibly due to the strong bias for the existence of $p = 0$ condensate. Evidently, $G$-state(SPB) does not describe the true $G$-state of a SIB. The analysis further concludes (Appendix 1) that the true $G$-state is described by $G$-state(PPB) having no particle with $p < h/2d$, which establishes the absence of $p = 0$ condensate in He-II. It is, obviously, not surprising that a number of recent studies of LHe-4 or smaller systems of $^4$He atoms (published over the last 15 years) report several interesting new results, viz. the loss of viscosity for the rotation of molecules embedded in microscopic clusters and droplets[15–18], linear Stark effect in roton mode observed in the absorption of microwaves[19–21], etc. which question the existence of the $p = 0$ condensate.

Neutron inelastic scattering experiments, supposed to render a direct proof for the existence of the $p = 0$ condensate, have been performed using neutron beams of different energies and their results are reviewed in several reports[3,11,22]. While different reports analysing these experiments conclude different values of $p = 0$ condensate ranging from 0% to about 20% (refs 3, 11, 22), a value around 10% has been accepted for the strong bias of people in favour of its existence, otherwise several experts raise doubts on its existence[23]. However, the said bias still exists (possibly for reasons similar to those counted by Hirsch[24] for the strong bias in favour of BCS theory of superconductivity). Consequently, new ideas and approaches to understand the LT behaviour of a SIB are not finding their due place, although a microscopic theory that explains the experimental properties of LHe-4 at quantitative scale is still awaited. In addition, we also lack right understanding of: (i) superfluidity, (ii) the true $G$-state and (iii) the real nature of BEC that exists in a SIB. Unaware of the role of a bias[24] (for conventional ideas/approaches) which works against the progress of new ideas/approaches, one of us used a new approach and concluded his microscopic theory[8] (without making any presumption of the existence of any condensate) that not only explains the properties of LHe-4 at quantitative scale but also concludes: (i) the absence of $p = 0$ condensate and (ii) true $G$-state which agrees with $G$-state(PPB) and related inferences by Jain[14] and (iii) the real nature of BEC that exists in He-II. The absence of the $p = 0$ condensate is also supported strongly[25,26] by the physical reality of the existence of electron bubble in LHe-4 (ref. 27) and the experimentally observed spectroscopy of embedded molecules[15–18].

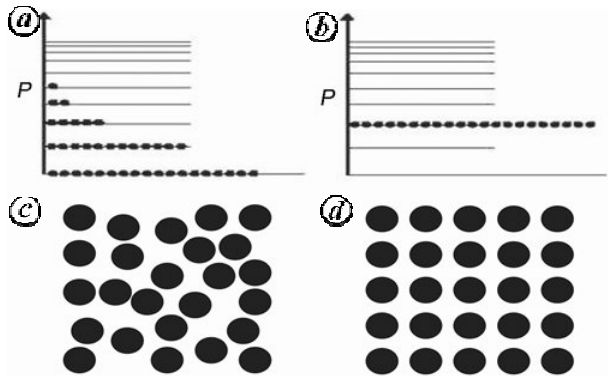

**Figure 1.** Schematic representation of momentum distribution of $N$ bosons in their ground state in accordance with: (**a**) Conventional microscopic theories based on Bogoliubov model[10] and (**b**) non-conventional microscopic theories concluded in Jain[8,14]; corresponding locations in normal space are depicted in **c** and **d** respectively.

Next we analyse the observed effect[19–21] not only to understand its origin but also to identify it as another experimental evidence for the absence of $p = 0$ condensate in He-II; to this effect we summarize relevant important aspects of a SIB and roton excitation in He-II in Appendix 1.

In what follows from Rybalko et al.[19–21] we note that:

(a) Microwave absorption in LHe-4 is found to peak at the frequency (corresponding to the energy of its well-known quantum quasi-particle excitation known as roton) which increases smoothly from $\approx 5.2$ K ($\equiv 125.0$ GHz) at $T = T_\lambda$ to $\approx 8.65$ K ($\equiv 180.3$ GHz) at $T = 0$, with its width decreasing from $\approx 400$ KHz at 2.2 K to $\approx 40$ KHz at 1.6 K.

(b) Under the influence of external electric field, $E_{ex}$, the said peak splits into two Stark components separated by

$$\Delta f = \frac{pE_{ex}}{h} \approx 2.8 \times 10^{-34} \frac{E_{ex}}{h}, \quad (1)$$

showing its linear dependence on $E_{ex}$ with $p \approx 2.8 \times 10^{-34}$ SI (here and henceforth SI refers to the SI units of the physical quantity) being the experimental value of the electric dipole moment, presumably, of a $^4$He atom.

(c) $p$ in eq. (1) (describing the experimental observation) remains constant for the entire range of experimental $E_{ex} = 0$ to $4.0 \times 10^5$ SI (refs 19–21) indicating that: (i) $E_{ex}$ has insignificant effect in inducing $p$ in spherically symmetric $^4$He atom; (ii) it is the internal electric field $E_{in}$ (seen by a $^4$He atom at its site) which induces $p$, and (iii) the strength of $E_{in}$ is much larger than the maximum $E_{ex}$ ($\approx 4 \times 10^5$ SI) used in the experiment.

(d) Using $p = \alpha E_{in}$ and atomic polarizability $\alpha = 0.1232$ cm$^3$/mol $= 2.1 \times 10^{-41}$ SI, we find

$$E_{in} \approx 1.3 \times 10^7 \text{ SI}. \quad (2)$$

The question about the origin of such a strong $E_{in}$ finds no answer from the our conventional understanding of He-II[4–6,16] summarized in Appendix 1 and represented by Figure 1 *a* and *c*; in what follows, atoms in He-II have relative motions and inter-particle collisions for which the positions and directions of their dipoles are bound to change randomly and this means that the net electric field at the site of every atom gets averaged out to zero.

In variance, a NCMT developed by Jain[8,14] concludes a different picture which is briefly discussed and summarized in Appendix 1 and depicted by Figure 1 *b* and *d*. Accordingly, all atoms not only constitute a close packed arrangement of wave packets (CPA-WP), but also assume a kind of collective binding[8] which provides certain amount of stability to CPA-WP against the thermal motions of the system and small energy perturbations such as flow of the fluid with velocity below a certain critical value, and the entire system behaves like a single macroscopic molecule. The CPA-WP allows atoms to move coherently in order of their locations and forbid them to have relative motions and mutual collisions. Naturally, if $^4$He atoms happen to have any electric dipole moment (possibly for the deformation of their electron density due to their mutual closeness), CPA-WP can allow their dipoles to align in a single direction, particularly when they are subjected to a $E_{ex}$ of even of $0^+$ (slightly above zero strength). If it is energetically favourable, unidirectional alignment of dipoles may also be possible in the absence of any $E_{ex}$, may be in the entire sample or over the scales of domains of the size of coherence length, $\xi$, estimated to be of the order of 100 Å (ref. 28). In view of the fact that the thermal motions of He-II represent a gas of non-interacting quantum quasi-particles which move in the system without disturbing the CPA-WP (Appendix 1), the collective binding of atoms with their CPA-WP can also add to the stability of the unidirectional alignment of dipoles against the thermal motions of He-II.

Here it is interesting to note that: (i) Rybalko et al.[19], on the basis of their experimental observations, believe that the relative motion of the normal and superfluid components is a result of internal electromagnetic forces related to the macroscopic quantum ordering of the system, and (ii) examining the possible reasons of the effect, Tomchenko[29] concludes that all atoms in He-II, acquiring small fluctuating dipole and multi-pole moments (oriented chaotically on the average) become partially ordered in the presence of a temperature or density gradient leading to volume polarization of He-II. In other words, Rybalko et al.[19] and Tomchenko[29] too believe in the ordering of atoms in terms of their positions and the direction of their dipoles as a necessary factor for the observation of the effect. The fact that the desired order is an obvious conclusion of NCMT[8], the possibility of existence of the $p = 0$ condensate is clearly ruled out. Evidently, the details of experimentally observed absorption not only supports NCMT[8] but also underline the absence of $p = 0$ condensate in superfluid $^4$He.

In what follows, one can use Feynman's relation[30]:

$$E_{in} = \frac{p}{\varepsilon_0} \frac{0.3812}{d^3}, \qquad (3)$$

developed for the electric field at the site of a dipole in an orderly simple cubic arrangement of very large number of dipoles oriented in a single direction. In eq. (3), $\varepsilon_0 = 8.854 \times 10^{-12}$ SI represents the dielectric permittivity of vacuum and $d$ the identically equal inter-dipole (atomic) distance which has a value of $3.57 \times 10^{-10}$ m for superfluid $^4$He. Using these values with $p = 2.8 \times 10^{-34}$ SI, we have $E_{in} = 2.6 \times 10^5$ SI which, however, is about two orders of magnitude smaller than $E_{in} \approx 10^7$ SI (eq. (2)) needed for producing $p \approx 10^{-34}$ SI units. Hence we examine whether this discrepancy arises from the following possibilities.

(i) Feynman's relation[30] assumes an arrangement of classically fixed dipoles with simple cubic structure while each dipole in He-II has position uncertainty of the order of $d$ itself, which suggests that $1/d^3$ in eq. (3) needs to be replaced by $\langle 1/r^3 \rangle$ (quantum mechanical average of $1/r^3$, where $r$ is the possible distance between two dipoles). However, the fact that two $^4$He atoms, due to their hard core nature, have zero probability to have a $r < 2.6$ Å (an approximate value of the hard core diameter of a $^4$He atom), indicates that $1/\langle r^3 \rangle$ cannot be $> 1/(2.6)^3$ Å$^{-3}$ which means that the replacement of $1/d^3$ by $\langle 1/r^3 \rangle$ can provide a maximum increase in $E_{in}$ by a factor of about $(3.57/2.6)^3 \approx 2.5$ (far shorter than the required factor of $\approx 100$). Evidently, the said replacement cannot be expected to render the desired result.

(ii) $^4$He atoms in the experimental resonant absorption cell interact more strongly with the metallic surface in comparison to their mutual interaction, and for this reason they are likely to get relatively more polarized, even in the absence of $E_{ex}$. This polarization could be the source of an additional electric field near the surface of the cell. If this renders the desired $E_{in}$, it would support the suggestion by Rybalko et al.[19–21] that the said absorption of microwave photons possibly occurs in atoms located near the said surface, because this possibility helps in understanding the conservation of linear momentum in the process of single photon absorption by a roton in He-II having very large momentum $Q \approx 1.93$ Å$^{-1}$, whereas a microwave photon has nearly zero momentum since the walls of the cell can take away the recoil momentum. We also note that for the same reasons Raman scattering, observed at twice the energy of the roton, is believed to create two rotons of equal and opposite $Q$[31]. However, the fact that $^4$He atoms in He-II form a CPA-WP which can easily absorb the said momentum of recoil since it allows particles to move coherently and a minimum number of such atoms falls around $(\xi/d)^3 \approx 20,000$, indicates that absorption can occur anywhere in the cell. Evidently, well-thought, additional experimental studies are needed to reach a definite conclusion.

(iii) A roton represents a collective excitation (as summarized in Appendix 1) of atoms. Assuming that it has about 100 atoms with dipole moment oriented orderly in one direction with a total dipole moment equal to $2.8 \times 10^{-34}$ SI, we find that per particle dipole moment is reduced to $2.8 \times 10^{-36}$ SI, for which the required $E_{in}$ has to be of order of $10^5$ SI. This seems to resolve the problem provided our assumption is true.

In conclusion, analysing the experimental observation of the Stark effect of a roton transition in He-II[19–21], we conclude: (i) The observation does not support CMT[4–6,10] for its conclusion that particles in the G-state of a SIB have random distribution of particles in momentum space and position space (Figure 1 *a* and *c*) and the presumed existence of $p = 0$ condensate because such a distribution does not have the required order. This is so because it violates an important law of nature that the true *G*-state of a physical system has to have minimum possible energy as established unequivocally by a brief analysis in Appendix 1 and a detailed study[14]. (ii) The observation supports an orderly arrangement of particles (CPA-WP) in position space (as concluded by a NCMT[8]) because it corresponds to minimum possible energy of the true *G*-state of a SIB and does not allow inter-particle collisions for which electric dipoles of $^4$He-atoms can have preferred orientation in specific direction, may be in the entire sample or in a domain of the size of $\xi$. As such the observation supports NCMT[8] and renders another experimental evidence against the presence of $p = 0$ condensate in He-II. (iii) A brief analysis of possible situations seems to indicate that the origin of $E_{in}$ and related aspects can be concluded only when we have the correct understanding of a roton. This would be possible only after comprehensive experimental and theoretical studies which conclude the correct description of a roton in the light of this analysis and without any bias for the existence of $p = 0$ condensate and related $N_p$.

Undoubtedly, superfluidity is related to current–current correlation and phase rigidity of collective motion of particles. However, CMT[4–6,10] does not explain how different atoms having different energy/momenta keep phase rigidity and coherence in their motion, or how $\approx 10\%$ $^4$He atoms in $p = 0$ condensate force the rest of the 90% to have current–current correlation with phase rigidity. This is particularly important since He-II at $T = 0$ is 100% superfluid. It is a well-established theoretical and experimental fact that waves (independent of their electromagnetic or de Broglie nature) of different $\lambda$ cannot have phase rigidity and coherence. Naturally, $^4$He-atoms in He-II having different momenta cannot be believed to have phase rigidity and coherence by violating this fact.

Since Landau phenomenology successfully explains superfluidity and related aspects of He-II, a viable microscopic theory should reveal clear reasons for its behaviour as a homogeneous mixture of two fluids of different properties. However, the basic aspects of CMT have serious difficulty in relating $p = 0$ condensate and superfluid density in any understandable manner. Consequently, several arguments are made to ease out this difficulty, viz. (i) Landau theory does not assume the existence of any condensate; (ii) there exists no conclusive proof that superfluidity necessarily requires any condensate; (iii) superfluid density and condensate fraction are not same, and (iv) BEC and superfluidity are two independent concepts[3] which seem to stand as the statements of facts but only in relation to CMT. However, CMT and its concluded $N_p$ have several questionable aspects (Appendix 1).

Similarly, as discussed by Reatto and Galli[32], description of roton dynamics concluded from CMT still remains in confusion. However, if we follow NCMT[8], description of the roton is lot more clear; accordingly, it is basically an excitation of single particle trapped in a cavity of size *d* which combines with collective motions of He-II for the strong inter-atomic momentum/energy correlations arising from CPA-WP of $^4$He atoms. However, we need to quantify the effective number of atoms that participate in the dynamics of roton, and we hope that future studies of LHe-4 would be able to reach a definite conclusion in this respect.

In Appendix 1, we not only present clear experimental evidence for the existence of CPA-WP in He-II, but also underline reasons for which this arrangement cannot be seen through diffraction of X-rays, neutrons, etc. The fact that the excitation spectrum of He-II shows only one branch of longitudinal acoustic phonons and no branches of transverse acoustic phonons, clearly concludes that $^4$He-atoms in He-II have vanishingly small shear forces, obviously, because of its fluidity and it is in this respect that CPA-WP differ from crystals which have reasonably strong shear forces. CPA-WP are obviously a fragile arrangement where particle positions have large uncertainties, as large as the inter-particle distance *d*. Naturally, this arrangement cannot be assigned a structure of well-defined symmetry. It is a simple arrangement where each particle keeps a distance *d* from its nearest neighbours, but without strict periodicity seen in crystals. It is not difficult to visualize such an arrangement. Finally, it is important to note that in accordance with Jain[8,14], particles form a CPA-WP in the *G*-state of every SIB, where number of particles *N* can be arbitrarily small or large. And for this reason superfluidity observed in bulk He-II as well as in microscopic systems (droplets, clusters, etc.) has a common origin and this fact itself is good evidence for the accuracy of Jain[8].

**Appendix 1.  Relevant important aspects of a SIB**

*G*-state(SPB) and $E_0$(SPB)

To a good approximation, particles in a fluid move freely over a surface of constant potential ($V = 0$ for gases and $V = -V_0$ for liquids) unless they collide with each other or with the walls of the container, and this remains true even for the *G*-state of LHe-4, obviously for its fluidity. Since $-V_0$ for any fluid is determined (independent of the motions of its particles) by density of particles and inter-particle interactions, the minimum of its total energy is basically determined by the minimum of its kinetic energy, which obviously depends on the momentum distribution $N_p$ of its particles. We note that *G*-state(SPB) concluded from refs 4–6 and 10) has a $N_p$ (Figure 1 *a*) where different number of atoms have different momenta (cf. Figure 1 *a*), $k_1$, $k_2$, $k_3$, ... (expressed in wavenumber), ranging from $k = 0$ to a large multiple (order of $N^{1/3}$) of $\pi/L$ (*L* is the size of the container). In the following we find whether this kind of distribution really corresponds to minimum possible energy as expected for the true *G*-state of any system.

The dynamics of two particles (say, *P*1 and *P*2 interacting through a two-body central force), moving with $\mathbf{k}_1$ and $\mathbf{k}_2$ in the laboratory frame, can always be described[8,14] in terms of their relative $\mathbf{k} = 2\mathbf{q} = \mathbf{k}_2 - \mathbf{k}_1$, and the centre of mass (CM) $\mathbf{K} = \mathbf{k}_1 + \mathbf{k}_2$ which are as independent as $\mathbf{k}_1$ and $\mathbf{k}_2$. We also have

$$\mathbf{k}_1 = -\mathbf{q} + \mathbf{K}/2, \quad (A1)$$

$$\mathbf{k}_2 = \mathbf{q} + \mathbf{K}/2. \quad (A2)$$

Since the CM motion represents a freely moving body of mass $2m$ because this motion does not encounter the inter-particle interaction, the *G*-state of *P*1 and *P*2 should invariably have $|\mathbf{K}| = 0$, which reduces corresponding energy to zero (the minimum energy of an interaction free motion). The fact that no two particles with different momenta have $\mathbf{K} = 0$ unless they have equal and opposite momenta, ($\mathbf{q}$ and $-\mathbf{q}$) and this is not true for all the ($N - 1$) pairs, a particle has with other $N - 1$ particles or a total of $N(N - 1)/2$ possible pairs that we can make in a system of $N$ particles. Evidently, *G*-state(SPB) does not have minimum possible energy as expected. Consequently, $N_p$ depicted in Figure 1 *a* does not represent the true *G*-state of a SIB.

### *G*-state(PPB), the true *G*-state

When *P*1 and *P*2 occupy their *G*-state, $\mathbf{K} = 0$ leaves $|\mathbf{k}_1| = q$ and $|\mathbf{k}_2| = q$ as their residual momentum, which is expected to have non-zero value (say, $q_0$) due to wave–particle duality. This should be true for all the $N - 1$ pairs involving one particle and in fact for a total of $N(N - 1)/2$ different pairs that we can count in the system. In order to find whether all particles would have identically equal $q_0$, we presume that one particle (say, *P*1) in the *G*-state has a $|\mathbf{q}| = |\mathbf{q}'|$ different from other particles having $|\mathbf{q}| = |\mathbf{q}_0|$. One immediately observes that all the $N - 1$ pairs involving *P*1 do not satisfy $\mathbf{K} = 0$ expected to hold for the true *G*-state, which means that all particles need to have equal $q_0$. Whereas a recent study by one of us[14] uses all such observations to conclude $q_0 = \pi/d$ which agrees exactly with his rigorous microscopic theory reported by Jain[8], here we obtain it from the simple facts: (i) a HC $^4$He atom does not share a volume of the order of $d^3$ with any other atom; (ii) a quantum particle manifests itself as WP of size $\lambda/2 = \pi/q$; (iii) for the fact that two HC particles do not overlap in position space, it is obvious that their representative WPs too do not overlap, and (iv) all particles having equal $q_0$ in the true *G*-state have equal $\lambda/2$. Evidently, two such WPs (HC particles) keep a distance $r \geq \lambda/2$, implying that particles having lowest possible $q = q_0$ or largest possible $\lambda/2$ ($= \pi/q_0$) can be equal to $d$; it cannot be larger than $d$ since the latter (decided independently by inter-particle interactions) is not expected to increase arbitrarily to accommodate WPs of size $\lambda/2 > d$. This renders $q_0 = \pi/d$. Thus the true *G*-state (i.e. *G*-state(PPB)) so concluded has the following details.

(1) Particles have identically equal zero-point momentum $q = q_0 = \pi/d$, corresponding zero-point energy $\varepsilon_0 = h^2/8md^2$ and WP size $\lambda/2 = d$. It is clear that each WP touches all other WPs in its neighbour indicating that particles constitute a CPA-WP with identically equal inter-particle distance $d$ between all neighbouring atoms and relative phase separation $\Delta\varphi = 2q_0d = 2\pi$ [equivalent to $\Delta\varphi = 2n\pi$ (with $n = 1, 2, 3, \ldots$), since as concluded in Jain[8] CPA-WP represent a 3D network of standing matter waves or vice versa]. It appears that each particle is trapped in a cavity of size $d$.

(2) Not even a single particle has an energy $< \varepsilon_0$ (or momentum $p < h/2d$), which means that a question of many particles having $p = 0$ (i.e. the existence of $p = 0$ condensate) does not arise and it is for this reason that the existence of $p = 0$ condensate in superfluid $^4$He has not been experimentally observed beyond a point of doubt.

### Real nature of BEC in a SIB

In what follows from eqs (A1) and (A2), each particle in a SIB is a part or a representative of a pair of particles moving with ($\mathbf{q}, -\mathbf{q}$) momenta with respect to its CM which moves with momentum $\mathbf{K}$ in the laboratory frame. Accordingly, it has two motions (which we call as *q*-motion and *K*-motion) of the pair it represents. While the state of all particles having $K = 0$ is reached only at $T = 0$, the onset of $K = 0$ occurs at a $T = T_c$ (say) at which all particles have reached a state of $q = q_0$ and they have no energy left to lose from *q*-motions. Obviously, the lower bound of $T_c$ can be fixed at $T = T_0$ (the $T$ equivalent of the lowest energy, $\varepsilon_0$, of *q*-motions) and the upper bound at $T \approx 2T_0$ by presuming that the *K*-motions of particles at $T = T_0$, also have an energy of the order of $\varepsilon_0$. The fact that the experimentally observed $T_\lambda = 2.17$ K for LHe-4 and the theoretical relation concluded by microscopic theory[8] fall at $\approx 1.5T_0$ (for LHe-4 $T_0 \approx 1.4$ K), provides strong experimental and microscopic foundation for concluding that particles in a SIB do have their BEC, but this differs from its conventional description in terms of $p = 0$ condensate. The real nature of BEC in a SIB is a condensation of particles as a part (or a representative) of a pair of particles in a single quantum state of $\mathbf{K} = 0$, $q = q_0$ and energy $\varepsilon_0$.

### Origin of two fluids

The origin of two-fluid nature of He-II lies with the simple fact that each particle has two independent motions (*q*-motion and *K*-motion) of momenta *q* and *K*. Since He-II at $T = 0$ is 100% superfluid, one is expected to find the origin of its superfluidity and related aspects as the intrinsic properties of its *G*-state(PPB) (where particles retain only *q*-motions as their zero-point motions). Interestingly, we really find that this state has: (i) zero entropy because all particles occupy single quantum state; (ii) zero viscosity because particles locked in CPA-WP cease to have relative motions (the major cause of viscosity), and (iii) coherence of particle motions because CPA-WP provide necessary phase rigidity by locking particles at $r = d$ and $\Delta\varphi = 2n\pi$ for which they can move coherently in order of their locations.

Further since CPA-WP get hardly disturbed with change in *T* from $T = 0$ to $T = T_\lambda$ (because as discussed above almost all particles over this range of *T* have $q = q_0$ as found in $T = 0$ state), the excitations of He-II such as phonons, maxons, etc. arise mainly from the *K*-motions and their correlations. Since K-motions are a kind of free motions, these excitations too have to be non-interacting quantum quasi-particles, which, obviously, form a kind of gas that exists every where in He-II, and this gas accounts for the entire entropy and viscosity of He-II.

Evidently, while $^4$He atoms frozen with $q = q_0$ represent the superfluid component of He-II, the gas of quantum quasi-particles (excitations) represents the normal fluid component which has all the properties exactly attributed by Landau[12].

### *G*-state(PPB) and experimental evidence

(I) Diffraction experiments using X-ray, neutron and electron beam as tools are expected to provide accurate information about the atomic arrangement in any system provided the high energy/momentum of these radiations does not damage/perturb this arrangement. However, CPA-WP are highly fragile. Although atoms cease to have their relative motions for their CPA-WP, they remain located on a flat potential surface for the fluidity of He-II and they are free to move coherently (all with same velocity keeping their relative positions, residual momentum and the relative phase positions fixed) in order of their locations on a line/plane or a closed path. In a sense atoms in He-II can slip on a line/plane with respect to those on neighbouring lines/planes, and this possibility is consistent with vanishingly small shear forces in He-II for its fluidity. Evidently, atoms in He-II are likely to have collective motions when they are hit by particles of high energy/momentum in a beam of the said radiations. Even the relative distance of particles is expected to change since it depends on the size of their WPs, which depend on their residual *q* which can have large fluctuations when He-II is exposed to the said radiations. This shows that diffraction tools are not suitable to get any reliable information about the CPA-WP and it is for this reason that CPA-WP could not be detected for so long.

(II) Other experiments that prove the CPA-WP can be identified from the three basic aspects for which they lock particles at distance *d*, relative momentum at $k = 2q_0 = 2\pi/d$ and relative phase position $\Delta\varphi = 2n\pi$ are as follows.

(1) The excitation spectrum $E(Q)$ of He-II matches closely with that predicted for a mono-atomic chain with atoms separated by *d* not only at low *Q* (observed for most liquids), but also at high $Q(> 2\pi/d)$. Since momentum and energy of an excitation at such a high *Q* can be attributed to the motion of a single particle for the fact that the excitation wavelength $\Lambda$ (< *d*, the space occupied by a single particle); this clearly proves that particles are arranged in order with a separation *d*.

(2) Landau two-fluid model which explains the properties of He-II to a good accuracy attributes zero entropy and zero viscosity to superfluid component of He-II and CPA-WP representing the superfluid component have zero viscosity because particles cease to have relative motions for their distance being locked at $r = d$ and zero entropy because all particles occupy single quantum state of $q = q_0$.

(3) Superfluid is observed to have coherent motion and vortices of quantum circulation and their possibility demands a configuration like CPA-WP where particles satisfy $\Delta\varphi = 2n\pi$.

Although the above listed experimental observations have been existing for many years, they were not analysed to see the existence of CPA-WP because it was never anticipated in the framework of CMT. Finally, as we find from Jain[8], many other aspects of superfluid such as $T^3$ dependence of specific heat, infinitely high thermal conductivity, etc. too support CPA-WP.

### Roton and its description

Assuming that the *G*-state of He-II is represented by *G*-state(SPB), a number of theoretical studies found different descriptions of the roton. As revealed recently by Reato and Galli[32], roton has a varying degree of: (i) single particle excitation, (ii) non-quantized smoke ring and (iii) collective excitations depending on its *Q* with respect to $Q_0$ (the wave vector of roton, minimum); almost a similar description was concluded by Feynman and Cohen[33]. However, based on the fact that the true *G*-state of He-II is synonymous with *G*-state(PPB), a CPA-WP, where each atom is identified with a particle trapped in a cavity of size *d* formed by neighbouring atoms, the roton at $Q = Q_0$ appears to represent an excited state of such a particle. The fact that the roton $Q_0$ and energy $E(Q_0)$ for He-II match closely[34] with $2q_0 \approx 2\pi/d$ (momentum of one of the possible quantum states of a trapped particle) and corresponding energy $\delta\varepsilon_0 = \eta^2/2m[(2q_0)^2 - q_0^2] = 3\varepsilon_0$ supports this possibility; as argued earlier, any change in the energy/momentum of one particle in CPA-WP shakes the entire system (or at least over a distance of the order of coherence length[28]) and such an excitation of a single particle gets naturally dressed with collective motions such as phonons, which is normally understood as a source of effective mass of a particle in an interacting environment.